\documentclass[11pt,a4paper]{article}
\usepackage{jheppub}
\usepackage[bbgreekl]{mathbbol}
\usepackage{tikz}
\usetikzlibrary {decorations.pathmorphing, decorations.pathreplacing, decorations.shapes,arrows}
\usepackage{physics}
\usepackage{nicematrix}
\usepackage{bm}
\usepackage{blkarray}
\usepackage[toc,page]{appendix}

\tikzset{decorate sep/.style 2 args=
{decorate,decoration={shape backgrounds,shape=circle,shape size=#1,shape sep=#2}}}

\newcommand{\ie}{i.e.\ }

\newcommand{\Remark}[1]{{\color{black}}}

\newcommand{\bbepsilon}{\bbespilon}
\newcommand\crule[3][black]{\textcolor{#1}{\rule{#2}{#3}}}

\newcommand{\Pf}{\operatorname{Pf}}

\DeclareMathOperator{\sgn}{sgn}
\DeclareMathOperator{\detp}{det^\prime}

\DeclareMathOperator{\col}{col}
\DeclareMathOperator{\row}{row}

\DeclareMathAlphabet{\mathsl}{OT1}{cmr}{m}{sl}
\SetMathAlphabet{\mathsl}{bold}{OT1}{cmr}{bx}{sl}
\newcommand{\dsa}{\mathsl{a}}

\makeatletter
\def\amsbb{\use@mathgroup \M@U \symAMSb}
\makeatother

\title{Scaffolding Residues in Yang-Mills-Scalar \`a la CHY}

\author[a]{Zurab Jashi,}
\author[a,b]{Jaroslav Scheinpflug,}
\author[a,c]{and Yale Yauk}
\affiliation[a]{Perimeter Institute for Theoretical Physics, \\ Waterloo, ON N2L 2Y5, Canada}
\affiliation[b]{Jefferson Physical Laboratory, Harvard University, \\ Cambridge, MA 02138 USA}
\affiliation[c]{Max-Planck-Institut f\"ur Quantenoptik,\\  85748 Garching, Germany}

\emailAdd{zjashi@perimeterinstitute.ca}
\emailAdd{jscheinpflug@g.harvard.edu}
\emailAdd{yale.yauk@mpq.mpg.de}

\abstract{Motivated by recent work by Arkani-Hamed et al.\@ \cite{Arkani-Hamed:2023jry}, we compute the ``scaffolding'' residue of $2n$-scalar Yang-Mills-Scalar amplitudes to obtain pure $n$-gluon amplitudes \`a la Cachazo-He-Yuan (CHY). In particular, we show how the Pfaffian of $\Psi$, which is a matrix rich in structure, emerges from that of the simple $A$ matrix. The same CHY computation straightforwardly produces $n$-graviton amplitudes from $2n$-scalar amplitudes in the Einstein-Maxwell-Scalar theory.
We also consider partial ``scaffolding'' residues, i.e., general multi-collinear limits and their interplay with color-dressed amplitudes. These notes are the result of a Perimeter Scholars International (PSI) Winter School Project supervised by Prof. Freddy Cachazo.}

\begin{document}
\maketitle
\flushbottom
\newpage

\section{Introduction and review of CHY formulas}
A standard quantum field theoretical treatment of an $n$-gluon scattering process quickly leads to large proliferation of terms. It turns out that, these amplitudes, particularly at tree-level, possess a surprising and beautiful analytical structure, which was first revealed by Parke and Taylor in their seminal calculation of $n$-gluon amplitudes \cite{Parke:1986gb}. Since then, incredible strides have been made to unveil the deep mathematical and analytical structures that underlie scattering amplitudes \cite{Witten:2003nn, Cachazo:2004kj, Britto:2004ap, Roiban:2004yf, Bjerrum-Bohr:2005xoa, Risager:2005vk, britto2005direct, Alex, Bern:2008qj, Arkani-Hamed:2008bsc, Arkani-Hamed:2010zjl, ArkaniHamed:2012nw, Arkani-Hamed:2013jha,  Arkani-Hamed:2014bca, Carrasco:2016ygv, Arkani-Hamed:2017mur, Mastrolia:2018uzb, Cachazo:2021wsz}, see \cite{ManganoReview, Dixon:1996wi, Cachazo:2005ga, Bern:2007dw, Benincasa:2013faa, Dixon:2013uaa, Elvang:2013cua, Mizera:2019ose, Travaglini:2022uwo, Badger:2023eqz} for reviews.

The Cachazo-He-Yuan formulation (CHY) \cite{CHY1, CHY2,CHY3,CHY4,CHY5,Dolan:2013isa} of tree-level scattering amplitudes connects the space of Mandelstam invariants $s_{ab}=(k_a+k_b)^2$ and the moduli space of punctured Riemann spheres, ${\cal M}_{0,n}$, via the scattering equations \cite{CHY2}. Scattering amplitudes of massless particles in a large variety of theories can be written as
\begin{equation}
\label{eq:genCHY}
A^{\rm massless}_{n} = \frac{1}{\mathrm{Vol}(\mathrm{SL}(2,\amsbb C))}\int\prod_{i=1}^n \dd{u_i} \prod_{i=1}^n\delta\bigg(\sum_{\substack{k=1 \\ k\neq i}}^n \frac{s_{ik}}{u_i- u_k}\bigg)\, \mathcal{I}_L\,\mathcal{I}_R\,,
\end{equation} 
where ${\cal I}_L$ and ${\cal I}_R$ are functions of the basic $\mathrm{SL}(2,\amsbb C )$ covariant factors $1/(u_i-u_j)$, and polynomials in $k_i\cdot k_j$, $k_i\cdot \epsilon_j$ and $\epsilon_i\cdot \epsilon_j$. This means that all singularities of the amplitude must be produced as a result of approaching a boundary of the moduli space, ${\cal M}_{0,n}$. 

The terms in the argument of the Dirac delta are the scattering equations -- there is one equation for each particle $i$. The functions ${\cal I}_L$ and ${\cal I}_R$ are usually called the ``left" and ``right" integrands. Each integrand carries the same $\mathrm{SL}(2,\amsbb{C})$ weight.

The first tree-level scattering amplitudes that were discovered to have a CHY formulation were those of gluons in pure $U(N)$ Yang-Mills in arbitrary space-time dimensions \cite{CHY2}. The CHY formula for a partial amplitude in the canonical ordering is given by
\begin{equation}
A^{\rm YM}_{n}(1,2,\ldots ,n) = \frac{1}{\mathrm{Vol}(\mathrm{SL}(2,\amsbb
C))}\int\prod_{i=1}^n \dd{u_i} \prod_{i=1}^n\delta\bigg(\sum_{\substack{k=1 \\ k\neq i}}^n \frac{s_{ik}}{u_i- u_k}\bigg)\mathrm{PT}(1,2,\ldots ,n)\,\Pf' \Psi_{2n}\,,
\end{equation} 
where the two integrands in \eqref{eq:genCHY} are
\begin{equation}
\label{eq:basicIntegrand}
{\cal I}_L^{\rm YM}:= \mathrm{PT}(1,2,\ldots ,n):=\frac{1}{(u_1-u_2)(u_2-u_3)\cdots (u_n-u_1)}\quad  {\rm and}\quad  {\cal I}_R^{\rm YM}:=\Pf' \Psi_{2n}.
\end{equation}
The first integrand is known as the Parke-Taylor factor \cite{Parke:1986gb}, while the second one is the Pfaffian of $\Psi_{2n}$, a $2n\times 2n$ matrix, given in block form by  
\begin{equation}
\label{eq:Psi_matrix}
\Psi_{2n} = \begin{bmatrix}
A & -C^\intercal \\
C & B
\end{bmatrix},
\end{equation}
where $A$ and $B$ are $n\times n$ antisymmetric matrices with off-diagonal components 
$$ A_{ab}:= \frac{k_a\cdot k_b}{u_a-u_b} \quad {\rm and} \quad B_{ab}:= \frac{\epsilon_a\cdot \epsilon_b}{u_a-u_b}.$$
The matrix $C$ has components designed to make gauge invariance manifest,
$$
C_{aa}:=-\sum_{\substack{b=1\\b\neq a}}^n C_{ab} \quad {\rm and}\quad C_{ab}:= \frac{\epsilon_a\cdot k_b}{u_a-u_b}.
$$
Gauge invariance, \ie the vanishing of the amplitude when any polarization vector $\epsilon_a$ is replaced by the momentum $k_a$ of the corresponding particle, manifests simply as a linear algebra property of Pfaffians. 

Soon after these formulations were discovered, various procedures were introduced to construct new CHY formulations out of known ones \cite{CHY5}. One such procedure is the compactification of amplitudes. Compactifying Yang-Mills amplitudes led to a theory known as Yang-Mills-Scalar (YMS). Higher dimensional gluons with polarization vectors $\epsilon_a$ entirely in the uncompactified dimensions remain as gluons while those $\epsilon_a$ entirely in the internal dimensions are scalars. Particularly simple is the CHY formulation of the pure scalar amplitude in YMS,
\begin{equation}
\label{eq:YMS}
A^{\rm YMS:scalar}_{n}(1,2,\ldots ,n) = \frac{1}{\mathrm{Vol}(\mathrm{SL}(2,\amsbb
C))}\int\prod_{i=1}^n \dd{u_i} \prod_{i=1}^n\delta\!\left(E_i\right)\,\mathrm{PT}(1,2,\ldots ,n)\,\Pf X\,\Pf' A\,.
\end{equation}
Here $E_i$ is a short-hand notation for the scattering equations, $A$ is the $n\times n$ matrix defined above while $X$ is the matrix obtained from $B$ by setting all $\epsilon_a\cdot \epsilon_b =1$.

In recent work \cite{Arkani-Hamed:2023jry}, Arkani-Hamed et al. introduced the ``scaffolding'' residue, which is a ``maximal" multi-collinear limit. Applying the scaffolding residue to an amplitude of $2n$ scalars in the YMS theory leads to an amplitude of $n$ gluons, where polarization vectors emerge from the momenta of the scalars and gauge invariance is geometrized. 

One of the objectives of this Perimeter Scholars International Winter School Project is the derivation of gluon amplitudes from scalar amplitudes using the scaffolding residue computed using the CHY formulation. As a by-product, the same CHY derivations directly connect $2n$ scalar amplitudes in the Einstein-Maxwell-Scalar (EMS) theory (obtained by compactification of Einstein's theory) to obtain $n$-graviton amplitudes. 

In this note we also consider general multi-collinear limits of YMS and extensions to full color-dressed amplitudes. 

This work is organized as follows: in section 2, we introduce the compactification procedure, which allows one to obtain the YMS amplitude~\eqref{eq:YMS}. We proceed to see how a YM amplitude emerges as the most singular term of the YMS amplitude under the scaffolding limit using the CHY formalism. In section 3, we extend the scaffolding analysis to Einstein gravity. Finally, in section 4, we present extensions of the scaffolding residue to possibly non-maximal multi-collinear limits. We show that color-dressed amplitudes also produce the expected singularities under the multi-collinear limits.

\section{From YM to YMS and back: scaffolding scalar amplitudes in YMS}
\label{sec:YMS} 

An intricate web of connections among theories of massless particles was constructed using various operations on the CHY formulation of three basic theories \cite{CHY2}. These are the biadjoint scalar, Yang-Mills, and Einstein Gravity. These are obtained by combining the two basic integrands $\mathrm{PT}(1,2,\ldots n)$ and $\Pf'\Psi_{2n}$ presented in \eqref{eq:basicIntegrand}.

The simplest operation on CHY integrands is compactification and when applied to the Yang-Mills formula, it leads to the CHY formulation of Yang-Mills-Scalar amplitudes. The procedure is to start with an amplitude of $n$ gluons in $D=d+1$ dimensions and set their momenta $\mathcal{K}_a$ and polarization vectors $\mathcal{E}_a$ to
\begin{align}
\begin{split}
\label{eq:compactification}
    \mathcal{K}_a & := (k_a,0), \quad \forall a \in \{ 1,2,\ldots ,n\} \\
    \mathcal{E}_a & := \begin{cases}
        (\epsilon_a,0), & \text{for a gluon}, \\
        (0,0,\ldots, 0,1), & \text{for a scalar}.
    \end{cases}
\end{split}
\end{align}
Here $k_a,\epsilon_a$ are $d$-dimensional momenta and polarization vectors respectively.

To substitute this into the expression for $\Pf'\Psi_{2n}$ it is convenient to use the indices $p,q,r$ for gluons, $i,j,k$ for scalars and $a,b,c$ when the identity of the particle is not important. Using this, the non-zero inner products in the matrix $\Psi_{2n}$ become
\begin{align}\nonumber 
    \frac{\mathcal{K}_a\cdot \mathcal{K}_b}{u_a-u_b} = &\frac{k_a\cdot k_b}{u_a-u_b}, \qquad \frac{\mathcal{E}_p\cdot \mathcal{E}_q}{u_p-u_q} = \frac{\epsilon_p\cdot \epsilon_q}{u_p-u_q}, \qquad \frac{\mathcal{E}_i\cdot \mathcal{E}_j}{u_i-u_j} = \frac{1}{u_i-u_j}\,, \\
    &\quad \frac{\mathcal{E}_p\cdot\mathcal{K}_q}{u_p-u_q}=\frac{\epsilon_p\cdot k_q}{u_p-u_q}, \quad \text{and} \quad \frac{\mathcal{E}_p\cdot\mathcal{K}_j}{u_p-u_j}=\frac{\epsilon_p\cdot k_j}{u_p-u_j}\,.
\end{align}

Substituting into the $\Psi_{2n}$ matrix with columns and rows organized so that the ``polarization" of scalars are last, one finds
\begin{equation}
\label{eq:block_compactification}
\Psi_{2n} = \begin{bmatrix}
A & -{\hat C}^\intercal & 0 \\
\hat C & {\hat B} & 0 \\
0 & 0 & X_s
\end{bmatrix},
\end{equation}
where $A$ is the original $n
\times n$ matrix $A$ while $\hat B$ only contains gluons. The matrix $X_s$ has components $1/(u_i-u_j)$ and only involves scalars. Clearly, $\Pf'\Psi_{2n}=\Pf'\widehat\Psi_{2n} \Pf X_s$.

Our main case of interest is the case when all particles are scalars and therefore $\Pf'\Psi_{2n}=\Pf'A\,\, \Pf X_s$. Since there are only scalars in the amplitude we drop the subscript and write $\Pf X_s$ as $\Pf X$.

Combining all the pieces we find the CHY formula for the tree-level amplitude of $n$ scalars,
\begin{equation}
\label{eq:YMS_amplitude}
    A^{\rm YMS:scalar}_{n}(1,2,\ldots ,n) = \frac{1}{\mathrm{Vol}(\mathrm{SL}(2,\amsbb C))}\int\prod_{i=1}^n \dd{u_i} \prod_{i=1}^n\delta\!\left(E_i\right)\,\mathrm{PT}(1,2,\ldots ,n)\,\Pf X\,\Pf' A\,.
\end{equation}

\subsection{Scaffolding from YMS to YM}
\label{sec:YMSTOYM}
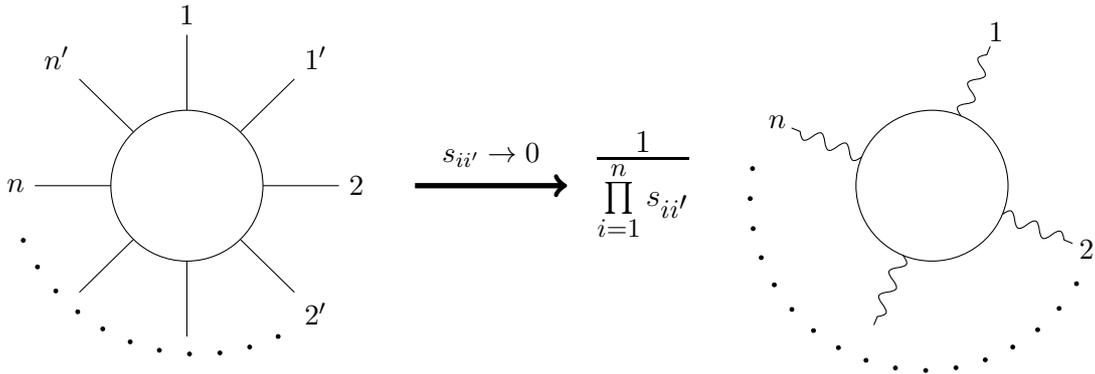
\begin{figure}
    \centering
    \begin{tikzpicture}
        \draw (-4, 0) circle (1);
        \draw (-4,1) -- (-4,2) node [above] {$1$};
        \draw (-3.293, 0.717) -- (-2.586, 1.414) node [above right] {$1'$};
        \draw (-3, 0) -- (-2, 0) node [right] {$2$};
        \draw (-3.293, -0.717) -- (-2.586, -1.414) node [below right] {$2'$};
        \draw[decorate sep={0.5mm}{4mm}, fill] (-2.8, -2) arc (-65:-160:2.5);
        \draw (-4,-1) -- (-4,-2);
        \draw (-4.707, -0.717) -- (-5.414,-1.414);
        \draw (-5, 0) -- (-6,0) node [left] {$n$};
        \draw (-4.707, 0.717) -- (-5.414, 1.414) node [above left] {$n'$};
        \draw[->, line width=2] (-1,0) -- (1,0);
        \node at (0,0.4) {$s_{ii'}\to 0$};
        \node at (2,0) {\LARGE$\frac{1}{\prod\limits_{i=1}^ns_{ii'}}$};
        \draw (5.8,0) circle (1);
        \draw[decorate, decoration={coil,aspect=0}] (6.18268, 0.92388) -- (6.56537, 1.84776);
        \node at (6.6419, 2.03253) {$1$};
        \draw[decorate, decoration={coil,aspect=0}] (6.72388, -0.382683) -- (7.64776, -0.765367);
        \node at (7.83253, -0.841904) {$2$};
        \draw[decorate, decoration={coil,aspect=0}] (5.41732, -0.92388) -- (5.03463, -1.84776);
        \draw[decorate sep={0.5mm}{4mm}, fill] (7.7, -1.3) arc (-30:-190:2.3);
        \draw[decorate, decoration={coil,aspect=0}] (4.87612, 0.382683) -- (3.95224, 0.765367);
        \node at (3.76747, 0.841904) {$n$};
    \end{tikzpicture}  
    \caption{Scaffolding of $2n$ scalars in the YMS theory to $n$ gluons in the YM theory.}
    \label{fig:scaffolding}
\end{figure}

In ref.~\cite{Arkani-Hamed:2023jry}, a new formulation for Yang-Mills scattering amplitudes at any loop order was proposed based on the same combinatorial techniques used for the $\Tr\phi^3$ theory. One of the key ingredients in making the connection was the use of a multi-collinear limit on Yang-Mills-Scalar amplitudes of $2n$ scalars to produce $n$-gluon amplitudes from ``scalar" data. This procedure was called the ``scaffolding residue", which we diagrammatically represent in figure~\ref{fig:scaffolding}.

Given that the CHY formalism straightforwardly connected YM amplitudes to that of YMS scalars, it is natural to see how the scaffolding residue is computed in the CHY formula for scalars \eqref{eq:YMS_amplitude} to produce that for gluons. At first sight, this seems an impossible task since the starting amplitude only contains the Pfaffian of a very simple matrix, the $A$ matrix, while the result should contain the Pfaffian of a matrix with a rich structure, the $\Psi$ matrix. In this section we show in detail how $\Pf'\Psi$ emerges from $\Pf'A$ in a beautiful manner.  

The scaffolding procedure takes a $2n$-scalar scattering amplitude in the YMS theory and produces an $n$-gluon scattering amplitude in the YM theory under the limit $s_{ii'}\to 0$ for each $1\leq i\leq n$. To this end, let us consider the YMS amplitude with $2n$ particles labelled $\{1,1',\dots, n,n'\}$: 
\begin{align}
\begin{split}
\label{eq:YMS_2n}
   & A^{\rm YMS:scalar}_{2n}(1,1',\ldots ,n, n') = \\ & \frac{1}{\mathrm{Vol}(\mathrm{SL}(2,\amsbb
C))}\int\prod_{i=1}^n \dd{u_i}\prod_{i'=1'}^{n'}\dd{u_{i'}} \prod_{i=1}^n\delta\!\left(E_i\right)\prod_{i'=1'}^{n'}\delta\!\left(E_{i'}\right)\,\mathrm{PT}(1,1',\ldots ,n,n')\,\Pf X\,\Pf' A\,.
\end{split}
\end{align}
The scaffolding limit $s_{ii'}\to 0$ can be parametrised by a variable $\tau$ and 
\begin{align}
\label{eq:change_variables}
    \begin{split}
        s_{ii'}&=\tau \hat{s}_{ii'}\,, \\
        u_{i'}&=u_i+\tau x_i\, .
    \end{split}
\end{align}
Here we have also assumed that $u_{i'}-u_i={\cal O}(\tau)$. In general the scattering equations for $2n$ particles have $(2n-3)!$ solutions. However, it is known that in a collinear limit, e.g. $s_{11'}\to 0$, the solutions split into ``singular" and ``regular". The terminology refers to the moduli space of punctured Riemann spheres. Solutions that go to the boundary of the moduli space are called singular while those that remain at generic point are called regular. In a single collinear limit, $s_{11'} ={\cal O}(\tau)$, singular solutions go as $u_{1'}-u_1={\cal O}(\tau)$. Moreover there are $((2n-1)-3)!$ such singular solutions. Since the scaffolding residue affects compatible channels, we expect that each $s_{ii'}\to 0$ limit can be described as a single collinear limit and hence $u_{i'}-u_i={\cal O}(\tau)$. In the end, after all $n$ collinear limits are implemented, there will only be $(n-3)!$ solutions.     

Let us introduce the commonly used shorthand notation,
\begin{equation}
    u_{ab} := u_a - u_b. 
\end{equation}
We apply the change of variables~\eqref{eq:change_variables} to the amplitude in eq.~\eqref{eq:YMS_2n}. Next we study each factor in the amplitude.

The simplest element to consider is the Parke-Taylor factor, which to leading order in $\tau$ becomes
\begin{equation}
\label{eq:PTtransform}
    \text{PT}(1,1',\dots,n,n')=\frac{1}{u_{11'}u_{1'2}u_{22'}\dots u_{nn'}u_{n'1}}=\frac{1}{(-1)^n\tau^n}\frac{1}{\prod_{i=1}^n x_i}\mathrm{PT}(1,\dots,n)\,,
\end{equation}
using the shorthand notation, $u_{ii'}= -\tau x_i$.


The next element is the Pfaffian of $X$. The explicit formula of the Pfaffian of a $2n\times2n$ antisymmetric matrix $M$ is given by
\begin{equation}
    \Pf M=\frac{1}{2^nn!}\sum_{\sigma\in S_{2n}}\sgn\sigma\prod_{i=1}^n M_{\sigma(2i-1),\sigma(2i)}\,.
\end{equation}
Recalling that the matrix elements of $X$ are given by $X_{ij}=\frac{1}{u_{ij}}$, the leading contribution to the Pfaffian of $X$ comes from the term where each factor in the product is in the form $u_{ii'}$, yielding, up to a sign,
\begin{equation}
\label{eq:Pfaffian_X_transform}
    \Pf X=\frac{1}{\tau^n}\frac{1}{\prod_{i=1}^nx_i}+O\left(\frac{1}{\tau^{n-1}}\right)\,.
\end{equation}

Next, consider the measure of integration. We choose to fix the $\mathrm{SL}(2,\amsbb C)$ redundancy by setting $u_1,u_2$ and $u_3$ to fixed values. Therefore, under the change of variables $u_{i'}= u_i+\tau x_i$, 
\begin{equation}
    \prod_{i=4}^n \dd{u_i}\prod_{i'=1}^{n'}\dd{u_{i'}}=\prod_{i=4}^n \dd{u_i}\tau^n\prod_{i=1}^n\dd{x_i}\,.
\end{equation}


Next, we have the scattering equations. To aid in the analysis, it is convenient to explicitly write down the scattering equations  $E_i=0$ and $E_{i'}=0$ to leading order in $\tau$,
\begin{align}
\label{eq:scatteringEq}
    E_i & =\sum_{\substack{j=1 \\ j\neq i}}^n\frac{s_{ij}}{u_{ij}}+\sum_{\substack{j'=1' \\ j'\neq i'}}^{n'}\frac{s_{ij'}}{u_{ij'}}+\frac{s_{ii'}}{u_{ii'}}=\sum_{\substack{j=1 \\ j\neq i}}^n\frac{s_{ij}}{u_{ij}}+\sum_{\substack{j'=1' \\ j'\neq i'}}^{n'}\frac{s_{ij'}}{u_{ij}}-\frac{\hat{s}_{ii'}}{x_i}\,,
\\
    E_{i'} & =\sum_{\substack{j=1 \\ j\neq i}}^n\frac{s_{i'j}}{u_{i'j}}+\sum_{\substack{j'=1' \\ j'\neq i'}}^{n'}\frac{s_{i'j'}}{u_{i'j'}}+\frac{s_{i'i}}{u_{i'i}}=\sum_{\substack{j=1 \\ j\neq i}}^n\frac{s_{i'j}}{u_{ij}}+\sum_{\substack{j'=1' \\ j'\neq i'}}^{n'}\frac{s_{i'j'}}{u_{ij}}+\frac{\hat{s}_{ii'}}{x_i}\,.
\end{align}

Adding the two equations, we get what will become a scattering equation for the $i^{\rm{th}}$-gluon, $\amsbb{E}_i=0$, 
\begin{equation}
    \amsbb{E}_i=E_i+E_{i'}=\sum_{\substack{j=1\\j\neq i}}^n\frac{s_{ij}+s_{ij'}+s_{i'j}+s_{i'j'}}{u_{ij}}\,.
\end{equation}
This is indeed the scattering equation we expect for Yang-Mills theory, if the momentum of the $i^{\rm th}$ gluon is defined as $\mathbb{k}_i:= k_i+k_{i'}$ and corresponding kinematic invariants $\mathbb{s}_{ij}:=(\mathbb{k}_i+\mathbb{k}_j)^2$, i.e.
\begin{align}
    \amsbb{E}_i&=\sum_{\substack{j=1 \\ j\neq i}}^n \frac{\mathbb{s}_{ij}}{u_{ij}}=\sum_{\substack{j=1 \\ j\neq i}}^n \frac{(\mathbb{k}_i+\mathbb{k}_j)^2}{u_{ij}} 
    =\sum_{\substack{j=1 \\ j\neq i}}^n\frac{(k_i+k_{i'}+k_j+k_{j'})^2}{u_{ij}} \nonumber \\
    &=\sum_{\substack{j=1 \\ j\neq i}}^n\frac{s_{ii'}+s_{ij}+s_{ij'}+s_{i'j}+s_{i'j'}+s_{jj'}}{u_{ij}} 
    = \sum_{\substack{j=1 \\ j\neq i}}^n\frac{s_{ij}+s_{ij'}+s_{i'j}+s_{i'j'}}{u_{ij}} + {\cal O}(\tau)\,.
\end{align}

Before considering the last element, \ie the Pfaffian of the $A$ matrix, let us collect the results so far using $\mathrm{SL}(2,\amsbb
C)$ to gauge fix $u_1,u_2$ and $u_3$, 
\begin{equation}
\label{eq:YMS_intermediate}
    A^{\mathrm{YMS}}_{11'\dots nn'}=\!\!\int\prod_{i=4}^n\dd{u_i}\prod_{i=1}^n\dd{x_i}\prod_{i=4}^n\delta(\amsbb{E}_i)\prod_{i'=1}^{n'}\delta(E_{i'})u_{12}^2u_{13}^2u_{23}^2\frac{1}{\tau^n}\frac{1}{\left(\prod_{i=1}^n x_i\right)^2}\,\mathrm{PT}(1\cdots n)\,\Pf'A\,.
\end{equation}

In this formula we interpret the equations $E_{i'}=0$ as equations for the $x_i$'s,
\begin{equation}
    E_{i'} =\sum_{\substack{j=1 \\ j\neq i}}^n\frac{s_{i'j}+s_{i'j'}}{u_{ij}}+\frac{\hat{s}_{ii'}}{x_i} = \sum_{\substack{j=1 \\ j\neq i}}^n\frac{2k_{i'}\cdot \mathbb{k}_j}{u_{ij}}+\frac{\hat{s}_{ii'}}{x_i} + {\cal O}(\tau)=0\,,
\end{equation} 
so that to leading order in $\tau$,
\begin{equation}
\label{eq:xForm}
    \frac{\hat{s}_{ii'}}{x_i} = -\sum_{\substack{j=1 \\ j\neq i}}^n\frac{2k_{i'}\cdot \mathbb{k}_j}{u_{ij}}\,.
\end{equation}
Interpreting $E_{i'}$ as a function of the $x_i$'s, we have the identity
\begin{equation}
\delta(E_{i'}) = \frac{(x^\circ_i)^2}{\hat{s}_{i i'}}\delta(x_i-x^\circ_i)\,,
\end{equation}
where $x^\circ_i$ solves \eqref{eq:xForm}. We carry out the integration over $x_i$ in eq.~\eqref{eq:YMS_intermediate}, leading to
\begin{equation}
\label{eq:preGluons}
    A^{\text{YMS}}_{11'\dots nn'}=\prod_{i=1}^n\frac{1}{s_{ii'}}\int\prod_{i=4}^n\dd{u_i}\prod_{i=4}^n\delta(\amsbb{E}_i)u_{12}^2u_{13}^2u_{23}^2\text{PT}(1,\dots,n)\Pf'A\, +\dots\,,
\end{equation}
where the ellipses indicate lower order terms in $\tau$. It is important to stress that the matrix $A$ still depends on $x_i$'s and therefore it is to be understood as evaluated on the solution $x_i^\circ$ to \eqref{eq:xForm}.

The scaffolding residue is defined as the coefficient of the most singular term in \eqref{eq:preGluons}. After inspecting the leading coefficient, it is clear that it has the correct structure of a $n$-particle amplitude with canonical color ordering. We are then left to prove that $\Pf'A = \Pf'\Psi$. 

There are several obstacles to overcome, the most obvious one is that $\Psi$ depends on the polarization vectors of the gluons while $A$ only depends on the momenta of the scalars. A precise map was proposed in \cite{Arkani-Hamed:2023jry} but we will not assume it and try to discover it by simply matching the momentum components of the matrix $\Psi$. 

\subsubsection{From the reduced Pfaffian of $A$ to that of $\Psi$}
\label{sec:Pfaffian_of_A}

The last ingredient to consider is the reduced Pfaffian of $A$, which is now to be evaluated at the solutions of the scattering equations \eqref{eq:xForm} for $x_i=x_i^\circ$, to show that
\begin{equation}
\label{eq:Pfaffian_A_transform}
    \Pf'A\rvert_{x_i=x_i^\circ}=\Pf'\Psi\,,
\end{equation}
where $\Psi$ is the antisymmetric $2n\times2n$ matrix given by
\begin{equation}
    \Psi=\begin{bmatrix}
        U & -W^\intercal \\ 
        W & V
    \end{bmatrix}\,,
\end{equation}
with the matrices of each block given by
\begin{align}
    U_{ab}&=\begin{cases}
        \frac{\mathbb{k}_a\vdot \mathbb{k}_b}{u_{ab}}, & a\neq b \\
        0, & a=b
    \end{cases}, \qquad V_{ab}=\begin{cases}
        \frac{\bbepsilon_a\vdot \bbepsilon_b}{u_{ab}}, & a\neq b \\
        0, & a=b
    \end{cases}, \qquad W_{ab}=\begin{cases}
        \frac{\mathbb{k}_a\vdot \bbepsilon_b}{u_{ab}}, & a\neq b \\
        -\sum_{\substack{c=1 \\ c\neq a}}^n\frac{\bbepsilon_a\vdot\mathbb{k}_c}{u_{ac}}, & a=b
    \end{cases}.
\end{align}
On the other hand, the matrix $A$ is given by its components 
\begin{equation}
    A_{IJ}=\begin{cases}
    \frac{k_I\cdot k_J}{u_{IJ}}, & I\neq J \\
    0, & I=J
\end{cases},
\end{equation}
where $I,J\in \{ 1,2,\dots,n,1',2',\dots,n'\}$. We choose to order the rows and columns by grouping all un-primed and primed labels together, \ie $(1,2,\dots,n,1',2', \ldots ,n')$. Note that this is different from that of the Parke-Taylor factor. Of course, the choice can only affect the result in an overall sign. 

The choice of ordering induces the following block structure,
\begin{equation}\label{towA}
A_{2n} = \begin{bmatrix}
{\widehat A}_n & -{\widehat C}^\intercal_n \\
{\widehat C}_n & {\widehat B}_n
\end{bmatrix},
\end{equation}
where 
\begin{equation}
{\widehat A}_n := \begin{bmatrix}
0 & \frac{k_1\cdot k_2}{u_1-u_2} & \cdots & \frac{k_1\cdot k_n}{u_1-u_n} \\
 \frac{k_2\cdot k_1}{u_2-u_1} & 0 & \cdots & \frac{k_2\cdot k_n}{u_2-u_n}  \\
\vdots & \vdots & \ddots & \vdots \\
 \frac{k_n\cdot k_1}{u_n-u_1} & \frac{k_n\cdot k_2}{u_n-u_2}  & \cdots & 0 \\
\end{bmatrix}, \quad 
{\widehat C}_n := \begin{bmatrix}
 \frac{\hat s_{1'1}}{2x_1} & \frac{k_{1'}\cdot k_{2}}{u_1-u_2} & \cdots & \frac{k_{1'}\cdot k_{n}}{u_1-u_n} \\
  \frac{k_{2'}\cdot k_{1}}{u_2-u_1} & \frac{\hat s_{2'2}}{2x_2} & \cdots & \frac{k_{2'}\cdot k_{n}}{u_2-u_n} \\
 \vdots & \vdots & \ddots & \vdots \\
 \frac{k_{n'}\cdot k_{1}}{u_n-u_1} & \frac{k_{n'}\cdot k_{2}}{u_n-u_2} & \cdots & \frac{\hat s_{n'n}}{2x_n} \\
\end{bmatrix},
\end{equation}

and 

\begin{equation}
{\widehat B}_n := \begin{bmatrix}
0 & \frac{k_{1'}\cdot k_{2'}}{u_1-u_2} & \cdots & \frac{k_{1'}\cdot k_{n'}}{u_1-u_n} \\
 \frac{k_{2'}\cdot k_{1'}}{u_2-u_1} & 0 & \cdots & \frac{k_{2'}\cdot k_{n'}}{u_2-u_n}  \\
\vdots & \vdots & \ddots & \vdots \\
 \frac{k_{n'}\cdot k_{1'}}{u_n-u_1} & \frac{k_{n'}\cdot k_{2'}}{u_n-u_2}  & \cdots & 0
\end{bmatrix}.
\end{equation}

Note that we have chosen to write the diagonal elements of the ${\widehat C}_n$ as
\begin{equation}
    \left( {\widehat C}_n \right)_{ii} =\frac{k_{i'}\cdot k_{i}}{u_{i'}-u_{i}} =  \frac{\hat s_{i'i}}{2x_i}, 
\end{equation}
by using that $u_{i'}-u_{i}=\tau x_i$, $s_{i'i}=2k_{i'}\cdot k_{i}$, and $s_{i'i}=\tau {\hat s}_{i'i}$.

Comparing the matrix ${\widehat A}_n$ with the top left block of the matrix $\Psi$ we notice that there is a mismatch that has to be fixed. While the entries of one matrix are $k_i\cdot k_j/u_{ij}$ the entries in the other are $\mathbb{k}_i\cdot \mathbb{k}_j/u_{ij}$. This is easily fixed by performing row and column operations on the matrix $A_{2n}$ in \eqref{towA}. 

We perform the following row and column operations to $A_{2n}$: 
$\col{i}\to\col{i}+\col{i'}$ followed by $\row{i}\to\row{i}+\row{i'}$, performed for each $1\leq i\leq n$.

This operation gives rise to a new matrix, 
\begin{equation}\label{finalA}
A^{\rm final}_{2n} =\begin{bmatrix}
\widehat{A}_n+\widehat{C}_n-\widehat{C}_n^\intercal+\widehat{B}_n & & &-\widehat{C}_n^\intercal+\widehat{B}_n \\
\widehat{C}_n+\widehat{B}_n & & & \widehat{B}_n
\end{bmatrix}:= \begin{bmatrix}
A_n & -{C}^\intercal_n \\
C_n & B_n
\end{bmatrix},
\end{equation}
where
\begin{equation}\label{finalAn}
A_n= \begin{bmatrix}
0 & \frac{k_1\cdot k_2+k_1\cdot k_{2'}+k_{1'}\cdot k_2+k_{1'}\cdot k_{2'}}{u_1-u_2} & \cdots & \frac{k_1\cdot k_n+k_1\cdot k_{n'}+k_{1'}\cdot k_n+k_{1'}\cdot k_{n'}}{u_1-u_n} \\
 \frac{k_1\cdot k_2+k_1\cdot k_{2'}+k_{1'}\cdot k_2+k_{1'}\cdot k_{2'}}{u_2-u_1} & 0 & \cdots & \frac{k_n\cdot k_2+k_n\cdot k_{2'}+k_{n'}\cdot k_2+k_{n'}\cdot k_{2'}}{u_2-u_n}  \\
\vdots & \vdots & \ddots & \vdots \\
\frac{k_1\cdot k_n+k_1\cdot k_{n'}+k_{1'}\cdot k_n+k_{1'}\cdot k_{n'}}{u_n-u_1} & \frac{k_n\cdot k_2+k_n\cdot k_{2'}+k_{n'}\cdot k_2+k_{n'}\cdot k_{2'}}{u_n-u_2}  & \cdots & 0 \\
\end{bmatrix},
\end{equation}
\begin{equation}
C_n= \begin{bmatrix}
 \frac{\hat s_{1'1}}{2x_1} & \frac{k_{1'}\cdot (k_{2}+k_{2'})}{u_1-u_2} & \cdots & \frac{k_{1'}\cdot (k_{n}+k_{n'})}{u_1-u_n} \\
  \frac{k_{2'}\cdot (k_{1}+k_{1'})}{u_2-u_1} & \frac{\hat s_{2'2}}{2x_2} & \cdots & \frac{k_{2'}\cdot (k_{n}+k_{n'})}{u_2-u_n} \\
 \vdots & \vdots & \ddots & \vdots \\
 \frac{k_{n'}\cdot (k_{1}+k_{1'})}{u_n-u_1} & \frac{k_{n'}\cdot (k_{2}+k_{2'})}{u_n-u_2} & \cdots & \frac{\hat s_{n'n}}{2x_n} \\
\end{bmatrix},
\end{equation}
and $B_n=\widehat{B}_n$.

The matrix $A_n$ in \eqref{finalAn} now has the correct structure since
\begin{equation}
    \frac{k_i\cdot k_j+k_i\cdot k_{j'}+k_{i'}\cdot k_j+k_{i'}\cdot k_{j'}}{u_i-u_j} = \frac{\mathbb{k}_i\cdot \mathbb{k}_j}{u_i-u_j} + {\cal O}(\tau).
\end{equation}
The matrices $B_n$ and $C_n$ become the correct matrices in $\Psi$ if the identification $ k_{i'} = \bbepsilon_{i}$ is made. Using this one finds that 
\begin{equation}
B_n = \begin{bmatrix}
 0 & \frac{\bbepsilon_{1}\cdot \bbepsilon_2}{u_1-u_2} & \cdots & \frac{\bbepsilon_{1}\cdot \bbepsilon_n}{u_1-u_n} \\
  \frac{\bbepsilon_{2}\cdot \bbepsilon_1}{u_2-u_1} & 0 & \cdots & \frac{\bbepsilon_{2}\cdot \bbepsilon_n}{u_2-u_n} \\
 \vdots & \vdots & \ddots & \vdots \\
 \frac{\bbepsilon_{n}\cdot \bbepsilon_1}{u_n-u_1} & \frac{\bbepsilon_{n}\cdot \bbepsilon_2}{u_n-u_2} & \cdots & 0 \\
\end{bmatrix}, \quad \text{and} \quad
C_n = \begin{bmatrix}
 \frac{\hat s_{1'1}}{2x_1} & \frac{\bbepsilon_{1}\cdot \mathbb{k}_2}{u_1-u_2} & \cdots & \frac{\bbepsilon_{1}\cdot \mathbb{k}_n}{u_1-u_n} \\
  \frac{\bbepsilon_{2}\cdot \mathbb{k}_1}{u_2-u_1} & \frac{\hat s_{2'2}}{2x_2} & \cdots & \frac{\bbepsilon_{2}\cdot \mathbb{k}_n}{u_2-u_n} \\
 \vdots & \vdots & \ddots & \vdots \\
 \frac{\bbepsilon_{n}\cdot \mathbb{k}_1}{u_n-u_1} & \frac{\bbepsilon_{n}\cdot \mathbb{k}_2}{u_n-u_2} & \cdots & \frac{\hat s_{n'n}}{2x_n} \\
\end{bmatrix},
\end{equation}

At this point, the only entries not written in terms of the gluon variables are the diagonal entries of $C_n$. However, using scattering equations in the form of \eqref{eq:xForm}, one has 
\begin{equation}
\label{eq:scattering_eq_diag_terms}
    \frac{\hat s_{i'i}}{2x_i} = -\sum_{\substack{j=1 \\ j\neq i}}^n\frac{\bbepsilon_{i}\cdot \mathbb{k}_j}{u_{ij}},
\end{equation}
which is precisely the correct $C_{ii}$ entry in the matrix $\Psi$. In fact, the diagonal elements in the $\Psi$ matrix have always been the most mysterious and it is very satisfying to find that they naturally emerge as a consequence of a scattering equation. 

The conclusion is that to leading order in $\tau$, $A^\mathrm{final}_{2n}=\Psi\implies \Pf'A\rvert_{x_i=x_i^\circ}=\Pf'\Psi$. This means that we obtain the expected result for the scaffolding residue, \ie
\begin{equation}
\label{YMStoYM} 
    A^{\text{YMS}}(1,1',\dots,n,n')=\prod_{i=1}^n\frac{1}{s_{ii'}}A^{\text{YM}}(1,\dots,n)\, +\ldots\,.
\end{equation}

\subsubsection{Explicit map and gauge transformations}
The computation of the scaffolding residue using the CHY formulation naturally led to the following identification
\begin{align}
\begin{split}
\label{eq:identification}
    \mathbb{k}_i & := k_i + k_{i'}, \\
    \bbepsilon_i & := k_{i'}.
\end{split}
\end{align}
While the map of the momenta is clear from the meaning of collinear limits, the identification of polarization vectors seems arbitrary.

To recover more general identifications, recall that the Pfaffian of $\Psi$ is invariant under elementary row and column operations. Consider the matrix $\tilde{\Psi}$, which is obtained from $\Psi$ via the transformation $\bbepsilon_a\vdot P\to \bbepsilon_a\vdot P+\alpha_a\mathbb{k}_a\vdot P$, where $P=\bbepsilon_i$ or $P=\mathbb{k}_i$. One can show that this transformation is an elementary column and row operation on $\Psi$, so the Pfaffian is invariant. These transformations are nothing but the standard gauge transformations Yang-Mills theory.

\begin{figure}
    \centering
    \begin{tikzpicture}
        \draw[->, dashed] (2.12132, -1.5) -- (1.5,-3);
        \draw[->, dashed] (1.5,-3) -- (0,-3.62132);
        \draw[->, dashed] (0,-3.62132) -- (-1.5,-3);
        \draw[->, dashed] (-1.5,-3) -- (-2.12132, -1.5);
        \draw[->, dashed] (-2.12132, -1.5) -- (-1.5,0);
        \draw[->, dashed] (-1.5,0) -- (0, 0.62132);
        \draw[->, dashed] (0, 0.62132) -- (1.5,0);
        \draw[->, dashed] (1.5,0) -- (2.12132, -1.5);
        \draw[->] (2.12132, -1.9) -- (1.5,-3);
        \draw[->] (1.5,-3) -- (-0.9,-3.62132);
        \draw[->] (-0.9,-3.62132) -- (-1.5,-3);
        \draw[->] (-1.5,-3) -- (-2.12132, -2.8);
        \draw[->] (-2.12132, -2.8) -- (-1.5,0);
        \draw[->] (-1.5,0) -- (-0.7, 0.62132);
        \draw[->] (-0.7, 0.62132) -- (1.5,0);
        \draw[->] (1.5,0) -- (2.12132, -1.9);
        \draw[->,>=stealth, blue, thick] (-1.5,-3) -- (-1.5,0);
        \draw[->,>=stealth, blue, thick] (-1.5,0) -- (1.5,0);
        \draw[->,>=stealth, blue, thick] (1.5,0) -- (1.5,-3);
        \draw[->,>=stealth, blue, thick] (1.5,-3) -- (-1.5,-3);
        \node[below] at (0,0) {\textcolor{blue}{$\mathbb{k}_1$}};
        \node[left] at (1.5,-1.5) {\textcolor{blue}{$\mathbb{k}_2$}};
        \node[above] at (0,-3) {\textcolor{blue}{$\mathbb{k}_3$}};
        \node[right] at (-1.5,-1.5) {\textcolor{blue}{$\mathbb{k}_4$}};
    \end{tikzpicture}
    \caption{Gauge invariance in the Yang-Mills-scalar theory is a geometric statement that the vertices between blue gluon momenta vectors $\mathbb{k}_i$ can be shifted parallel to $\mathbb{k}_i$. Here we consider a scattering with $n=4$ gluons, with the dashed and solid black arrows corresponding to the momenta of scalar particles. Both configurations give rise to the same scattering amplitude due to gauge invariance.}
    \label{fig:gauge}
\end{figure}
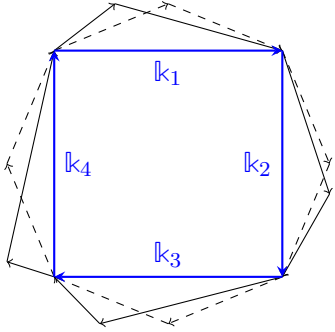

The corresponding gauge transformations in the scalar language are given by
\begin{align}
    \begin{split}
        k_i\to k_i-\alpha_i(k_i+k_{i'})\,, \\
        k_i'\to k_i'+\alpha_i(k_i+k_{i'})\, ,
    \end{split}
\end{align}
which are the most general transformations that leave the sum $\mathbb{k}_i=k_i+k_{i'}$ invariant and transform $\bbepsilon_i\to \bbepsilon_i+ \alpha_i \mathbb{k}_i$. This is the geometrization of gauge invariance observed in ref. \cite{Arkani-Hamed:2023jry}. Having a color ordering motivates the construction of a closed polygon with edges the momenta of the particles, $\{ k_1, k_{1'}, \dots, k_n, k_{n'} \}$. The polygon closes due to momentum conservation. Gauge invariance is then a statement that every vertex between momentum vectors $\{k_i,k_{i'}\}$ can be arbitrarily shifted in the direction of $k_i+k_{i'}$, while keeping the vertices between $\{k_{(i-1)'},k_{i}\}$ and $\{k_{i'},k_{i+1}\}$ fixed -- see figure~\ref{fig:gauge} for a visualization. This shows how the scalars know about gauge invariance of Yang-Mills theory, even though they do not carry any polarizations. 

\section{From GR to EMS and back: scaffolding scalar amplitudes in EMS}
In this section we repeat the analysis carried out above for Yang-Mills to Einstein gravity. The scattering amplitudes of the Einstein-Maxwell-Scalar (EMS) theory can be obtained using the CHY formalism by choosing the integrands to be $\mathcal{I}_L=\mathcal{I}_R=\Pf'{\Psi_{2n}}$, with the same matrix $\Psi_{2n}$ as defined in \eqref{eq:Psi_matrix} and applying the compatification procedure. Note that the polarization vector for both $\mathcal{I}_L$ and $\mathcal{I}_R$ are the same for a theory of pure gravity \cite{CHY1}. In the discussion following the compactification equations \eqref{eq:compactification}, we showed that for a scattering amplitude consisting entirely of scalars, $\Pf'\Psi_{2n}$ factorizes and becomes $\Pf X\Pf'A$. Thus we have that the pure gravity integrand $(\Pf'{\Psi_{2n}})^2$ becomes\footnote{Recall that $\Pf'A:=1/u_{ij}\Pf A_{ij}^{ij}$, where $A_{ij}^{ij}$ is the submatrix of $A$ obtained by deleting the $i^{\rm th}$ and $j^{\rm th}$ rows and columns, while $\detp A:=1/(u_{ij}u_{kl})\det A_{ij}^{kl}$. It is crucial that both $\Pf'A$ and $\detp A$ do not depend (up to a sign) on the choices of rows and columns that are removed.} $\det X\detp A$, leading to the tree-level scattering amplitude for $n$ EMS scalars
\begin{equation}
    A_n^{\mathrm{EMS:\, scalar}}(1,2,\dots,n)=\frac{1}{\mathrm{Vol}(\mathrm{SL}(2,\amsbb C))}\int\prod_{i=1}^n \dd{u_i} \prod_{i=1}^n\delta\!\left(E_i\right)\det X\detp A\,.
\end{equation}

\subsection{Scaffolding from EMS to EM}
We begin with the EMS amplitude for $2n$ scalars labelled $\{1,1',\dots,n,n'\}$:
\begin{align}
\begin{split}
\label{eq:EMSmain}
   & A^{\rm EMS:\, scalar}_{2n}(1,1',\ldots ,n, n') = \\ & \frac{1}{\mathrm{Vol}(\mathrm{SL}(2,\amsbb
C))}\int\prod_{i=1}^n \dd{u_i}\prod_{i'=1'}^{n'}\dd{u_{i'}} \prod_{i=1}^n\delta\!\left(E_i\right)\prod_{i'=1'}^{n'}\delta\!\left(E_{i'}\right)\,\det X\detp A\,.
\end{split}
\end{align}
The scaffolding limit is parametrized by the same variables as in eq.~\eqref{eq:change_variables}, which we rewrite here for convenience:
\begin{align}
    \begin{split}
        s_{ii'}&=\tau \hat{s}_{ii'}\,, \\
        u_{i'}&=u_i+\tau x_i\, .
    \end{split}
\end{align}
Again, we must study each factor in the amplitude and see how they change under the scaffolding limit. Actually, all of the hard work has been done in the previous section. Owing to the identity $(\Pf{X})^2=\det X$ for an antisymmetric matrix $X$, eq.~\eqref{eq:Pfaffian_X_transform} tells us that 
\begin{equation}
    \det X=\frac{1}{\tau^{2n}}\frac{1}{(\sum_{i=1}^n x_i)^2}+O\left(\frac{1}{\tau^{2n-1}}\right)\,,
\end{equation}
and eq.~\eqref{eq:Pfaffian_A_transform} says
\begin{equation}
    \detp A\rvert_{x_i=x_i^\circ}=(\Pf'\Psi)^2\,,
\end{equation}
with the correct momentum and polarization identifications given by eq.~\eqref{eq:identification}. The measure of integration and scattering equations transform as before giving us the scaffolding residue
\begin{equation}
    A^{\mathrm{EMS:\, scalar}}(1,1',\dots,n,n')=\prod_{i=1}^n\frac{1}{s_{ii'}} A^{\mathrm{EMS:\, graviton}}(1,2,\dots,n)+\dots\,,
\end{equation}
where
\begin{equation}
    A^{\mathrm{EMS:\, graviton}}(1,2,\dots,n)=\int\prod_{i=4}^n\dd{u_i}\prod_{i=4}^n\delta(\amsbb{E}_i)u_{12}^2u_{13}^2u_{23}^2(\Pf'\Psi)^2\,
\end{equation}
is the standard CHY formula for the $n$-graviton amplitude in pure Einstein gravity. 

\section{Multi-collinear limits and color-dressed amplitudes}
In this section, we generalize the discussion to possibly non-maximal multi-collinear limits. To be precise, a multi-collinear limit is a limit in which we take $s_{i i'} \to 0$ for some sub-collection of pairs $(1,1'), \ldots, (m, m')$, which does not necessarily exhaust all possible pairs of consecutive momenta \ie some of the external momenta remain unpaired. In this language, we understand the scaffolding limit as a particular case in which the pairings outlined above are perfect matchings, which was previously described as ``maximal''.

\subsection{Scalars in the YMS theory}
To study the multi-collinear limit for the YMS theory, we shall consider the case where, of the $n$ particles in the partial amplitude, the first $m<n$ particles are split into pairs $(i,i')$ with momenta $(k_i,k_{i'})$, where $1\leq i\leq m$. We perform a multi-collinear limit on the $m$ pairs of particles and leave the remaining $n-m$ particles untouched and show that the resulting amplitude is one of $m$ gluons and $n-m$ scalars. This is essentially a ``partial" scaffolding procedure. The partial amplitude~\eqref{eq:YMS_amplitude} of $n+m$ scalar particles labelled $(1,1',\dots,m'm,m+1,\dots,n)$ is our starting point:
\begin{align}
    \begin{split}
    & A_{n+m}^{\mathrm{YMS:scalar}}(1,1',\dots,m,m',m+1,\dots,n)=\\
    & \frac{1}{\mathrm{Vol}(\mathrm{SL}(2,\amsbb{C}))}\int \prod_{i=1}^m \mathfrak{d}u_i\prod_{j=1}^m \mathfrak{d}u_{j'}\!\!\! \prod_{a=m+1}^n \!\! \mathfrak{d}u_a \, \mathrm{PT}(1,1',\dots,m,m',\dots,n) \Pf{X} \Pf'{A}\,,
    \end{split}
\end{align}
where we introduced the compact notation $\mathfrak{d}u_i:=\dd{u_i}\delta(E_i)$ containing the measure and the scattering equation. We shall adhere to the convention where indices $1\leq i,j\leq m$ refer to the scaffolded particles, and the indices $m+1\leq a,b\leq n$ refer to unscaffolded scalar particles. To take the multi-collinear limit $s_{i i'} \to 0$ for $1\leq i\leq m$, we first need to revisit the analysis of scattering equations performed in subsection \ref{sec:YMSTOYM}, since extra terms associated to the index $a$ appear.

Using the change of variables in eq.~\eqref{eq:change_variables} as before, the deformed scattering equations read
\begin{align}
    0&=E_i=\sum\limits_{a=m+1}^{n} \frac{s_{i a}}{u_{ia}} + \sum_{\substack{j=1 \\ j\neq i}}^m\frac{s_{ij}+s_{i j'}}{u_{ij}}-\frac{\hat{s}_{ii'}}{x_i}\,, \\
    0&=E_{i'}= \sum\limits_{a=m+1}^{n} \frac{s_{i' a}}{u_{i'a}} + \sum_{\substack{j=1 \\ j\neq i}}^m\frac{s_{i'j}+s_{i'j'}}{u_{ij}}+\frac{\hat{s}_{ii'}}{x_i}\,, \\
    0&=E_{a}= \sum_{\substack{b=m+1 \\ b \neq a}}^n\frac{s_{a b}}{u_{a b}} + \sum_{\substack{j=1}}^m\frac{s_{a j}+s_{a j'}}{u_{a j}}\,.
\end{align} 
Adding the first two equations, we get that
\begin{equation}
    0=\amsbb{E}_i:=E_i+E_{i'}=\sum_{\substack{j=1\\j\neq i}}^m\frac{s_{ij}+s_{ij'}+s_{i'j}+s_{i'j'}}{u_{ij}} + \sum\limits_{a=m+1}^n \frac{s_{i a} + s_{i'a}}{u_{i a}}\,.
\end{equation}
Defining $\mathbb{k}_i := k_i + k_{i'}$ and $\mathbb{k}_a := k_a$ and assembling them into a single $\mathbb{k}_I$ where the index $I$ runs over both $i$ and $a$, one can form the corresponding kinematic invariant $\mathbb{s}_{IJ}$. Then we get
\begin{align}
    0&=\amsbb{E}_i= \sum_{\substack{j=1\\j\neq i}}^m\frac{\mathbb{s}_{i j}}{u_{ij}} + \sum\limits_{a=m+1}^n \frac{\mathbb{s}_{i a}}{u_{i a}} = \sum_{\substack{J=1 \\ J\neq i}}^n \frac{\mathbb{s}_{i J}}{u_{i J}}\,,  \\
    0&=E_{a}= \sum_{\substack{b=m+1 \\ b \neq a}}^n\frac{\mathbb{s}_{a b}}{u_{a b}} + \sum_{\substack{j=1}}^m\frac{\mathbb{s}_{a j}}{u_{a j}}\, = \sum_{\substack{I=1 \\ I \neq a}}^n \frac{\mathbb{s}_{a I}}{u_{a I}}\,.
\end{align} 
These are the expected scattering equations from an interaction of $n$ particles. Thus the scattering equations continue to behave nicely under the multi-collinear limit.

Next, we examine the Parke-Taylor factor. Following the logic of eq.~\eqref{eq:PTtransform}, we have 
\begin{equation}
    \mathrm{PT}(1,1',\dots, m, m', m+1, \dots, n) = \biggr(\prod\limits_{i=1}^m \frac{1}{\tau x_i} \biggr) \mathrm{PT}(1,\dots,n)\,.    
\end{equation}

Lastly, each primed measure of integration yields a power of $\tau$ under the change of variables $u_{j'}=u_j+\tau x_i$:
\begin{equation}
    \prod_{j=1}^m \mathfrak{d}u_{j'}=\tau^m\prod_{j=1}^m\bigg(\!\dd{x_j}\delta(E_{j'})\!\bigg)\,.
\end{equation}

It remains to study the transformation of $\Pf(X)$ and $\Pf'(A)$. Let $I,J$ be indices running over all particle labels $i$ and $a$. Then the matrices have the explicit form
\begin{equation}
    X_{IJ}=\begin{cases} \frac{1}{u_{IJ}}, & I\neq J \\
    0, & I=J
    \end{cases}\,,
    \qquad A_{IJ}=\begin{cases}
        \frac{k_I\cdot k_J}{u_{IJ}}, & I\neq J \\
        0, & I=J
    \end{cases}\,.
\end{equation}
In this basis, the matrix $X$ has an evocative block form
\begin{equation}
    X=\begin{bmatrix}
        M & Q \\
        -Q^\intercal & N
    \end{bmatrix},
\end{equation}
where $M=O(\tau^{-1})$ is a $2m\times 2m$ matrix, $Q=O(1)$ is $2m\times (n-m)$, and $N=O(1)$ is $(n-m)\times(n-m)$. We make the observation that $N$ is precisely the `$X$' matrix for the scalar particles whose momenta are not collinearized. Provided $M$ is invertible, we can write
\begin{equation}
    \Pf(X)=\Pf(M)\Pf(N+Q^\intercal M^{-1}Q)=\frac{1}{\tau^m\prod_{i=1}^m x_i}\Pf(N)+O\left(\frac{1}{\tau^{m-1}}\right)\,.
\end{equation}

Substituting the expressions for the factors we have studied to this point and gauge fixing $u_1,u_2$ and $u_3$, we have
\begin{align}
    \begin{split}
    & A_{n+m}^{\mathrm{YMS:scalar}}(1,1',\dots,m,m',m+1,\dots,n)=\\
    & \prod\limits_{k=1}^m \frac{1}{\tau x_k^2} \int \prod_{i=4}^m\dd{u_i}\delta(\amsbb{E}_i)\prod_{j=1}^m \dd{x_j}\delta(E_{j'})\!\! \prod_{a=m+1}^n \!\! \mathfrak{d}u_a \,(u_{12}u_{23}u_{31})^2\, \mathrm{PT}(1,\dots,n) \Pf{N} \Pf'{A}\,.
    \end{split}
\end{align}

Performing the integration over primed variables exactly as before in eq.~\eqref{eq:preGluons}, we get 
\begin{align}
    \begin{split}
    & A_{n+m}^{\mathrm{YMS:scalar}}(1,1',\dots,m,m',m+1,\dots,n)=\\
    & \biggr(\prod\limits_{k=1}^m \frac{1}{s_{kk'}} \biggr)\int \prod_{i=1}^m\bigg(\!\dd{u_i}\delta(\amsbb{E}_i)\!\bigg)\!\!\! \prod_{a=m+1}^n \!\! \mathfrak{d}u_a \,(u_{12}u_{23}u_{31})^2\,\mathrm{PT}(1,\dots,n) \Pf{N} \Pf'{A}\,,
    \end{split}
\end{align}
The $x_j$-dependence of $\Pf'{A}$ is understood to be evaluated to at the solutions $x_j^\circ$ of the primed scattering equations.

Now consider the matrix $A$, which we choose to represent in the basis $(1,\dots,m$,$1'$, $\dots$,$m'$,$m+1,\dots,n)$. It has the block form
\begin{equation}
    A_{(n+m)\times(n+m)}=\begin{bmatrix}
        \bm{\mathsf{A}}_{m\times m} & -\bm{\mathsf{C}}^\intercal_{m\times m} & -\bm{\mathsf{D}}^\intercal_{m\times (n-m)} \\
        \bm{\mathsf{C}}_{m\times m} & \bm{\mathsf{B}}_{m\times m} & -\bm{\mathsf{E}}^\intercal_{m\times (n-m)} \\
        \bm{\mathsf{D}}_{(n-m)\times m} & \bm{\mathsf{E}}_{(n-m)\times m} & \bm{\mathsf{F}}_{(n-m)\times (n-m)}
    \end{bmatrix},
\end{equation}
with the subscripts denoting the dimension of the block. Explicitly, we have
\begin{equation}
\begin{split}
    \bm{\mathsf{A}} & = \begin{bmatrix}
0 & \frac{k_1\cdot k_2}{u_{12}} & \cdots & \frac{k_1\cdot k_m}{u_{1m}} \\
 \frac{k_2\cdot k_1}{u_{21}} & 0 & \cdots & \frac{k_2\cdot k_m}{u_{2m}}  \\
\vdots & \vdots & \ddots & \vdots \\
 \frac{k_m\cdot k_1}{u_{m1}} & \frac{k_n\cdot k_2}{u_{m2}}  & \cdots & 0 \\
\end{bmatrix}, \\
\bm{\mathsf{B}} &= \begin{bmatrix}
0 & \frac{k_{1'}\cdot k_{2'}}{u_{12}} & \cdots & \frac{k_{1'}\cdot k_{m'}}{u_{1m}} \\
 \frac{k_{2'}\cdot k_{1'}}{u_{21}} & 0 & \cdots & \frac{k_{2'}\cdot k_{m'}}{u_{2m}}  \\
\vdots & \vdots & \ddots & \vdots \\
 \frac{k_{m'}\cdot k_{1'}}{u_{m1}} & \frac{k_{m'}\cdot k_{2'}}{u_{m2}}  & \cdots & 0
\end{bmatrix}, \\
\bm{\mathsf{E}} &= \begin{bmatrix}
 \frac{k_{m+1}\cdot k_{1'}}{u_{(m+1)1}} & \frac{k_{m+1}\cdot k_{2'}}{u_{(m+1)2}} & \cdots & \frac{k_{m+1}\cdot k_{m'}}{u_{(m+1)m}} \\
  \frac{k_{m+2}\cdot k_{1'}}{u_{(m+2)1}} & \frac{k_{m+2}\cdot k_{2'}}{u_{(m+2)2}} & \cdots & \frac{k_{m+2}\cdot k_{m'}}{u_{(m+2)m}} \\
 \vdots & \vdots & \ddots & \vdots \\
 \frac{k_{n}\cdot k_{1'}}{u_{n1}} & \frac{k_{n}\cdot k_{2'}}{u_{n2}} & \cdots & \frac{k_{n}\cdot k_{m'}}{u_{nm}} \\
\end{bmatrix},
\end{split}
\qquad 
\begin{split}
    \bm{\mathsf{C}}&= \begin{bmatrix}
 \frac{\hat s_{1'1}}{x_1} & \frac{k_{1'}\cdot k_{2}}{u_{12}} & \cdots & \frac{k_{1'}\cdot k_{m}}{u_{1m}} \\
  \frac{k_{2'}\cdot k_{1}}{u_{21}} & \frac{\hat s_{2'2}}{x_2} & \cdots & \frac{k_{2'}\cdot k_{m}}{u_{2m}} \\
 \vdots & \vdots & \ddots & \vdots \\
 \frac{k_{m'}\cdot k_{1}}{u_{m1}} & \frac{k_{m'}\cdot k_{2}}{u_{m2}} & \cdots & \frac{\hat s_{m'm}}{x_m} \\
\end{bmatrix}, \\
\bm{\mathsf{D}} &= \begin{bmatrix}
 \frac{k_{m+1}\cdot k_{1}}{u_{(m+1)1}} & \frac{k_{m+1}\cdot k_{2}}{u_{(m+1)2}} & \cdots & \frac{k_{m+1}\cdot k_{m}}{u_{(m+1)m}} \\
  \frac{k_{m+2}\cdot k_{1}}{u_{(m+2)1}} & \frac{k_{m+2}\cdot k_{2}}{u_{(m+2)2}} & \cdots & \frac{k_{m+2}\cdot k_{m}}{u_{(m+2)m}} \\
 \vdots & \vdots & \ddots & \vdots \\
 \frac{k_{n}\cdot k_{1}}{u_{n1}} & \frac{k_{n}\cdot k_{2}}{u_{n2}} & \cdots & \frac{k_{n}\cdot k_{m}}{u_{nm}} \\
\end{bmatrix}, \\
\bm{\mathsf{F}} &= \begin{bmatrix}
0 & \frac{k_{m+1}\cdot k_{m+2}}{u_{(m+1)(m+2)}} & \cdots & \frac{k_{m+1}\cdot k_{n}}{u_{(m+1)n}} \\
 \frac{k_{m+2}\cdot k_{m+1}}{u_{(m+2)(m+1)}} & 0 & \cdots & \frac{k_{m+2}\cdot k_{n}}{u_{(m+2)n}}  \\
\vdots & \vdots & \ddots & \vdots \\
 \frac{k_{n}\cdot k_{m+1}}{u_{n(m+1)}} & \frac{k_{n}\cdot k_{m+2}}{u_{n(m+2)}}  & \cdots & 0
\end{bmatrix}.
\end{split}
\end{equation}
Just as in section~\ref{sec:Pfaffian_of_A}, we perform column and row operations $\col{i}\to\col{i}+\col{i'}$ and $\row{i}\to\row{i}+\row{i'}$, for each $1\leq i\leq m$. After this, we additionally perform column and row swaps $\col{i'}\leftrightarrow\col{a}$ and $\row{i'}\leftrightarrow\row{a}$, \ie swapping the last two block columns and rows. With the proper choice of momenta and polarizations $\mathbb{k}_i:=k_i+k_{i'}$, $\mathbb{k}_a:=k_a$ and $\bbepsilon_i:=k_{i'}$, this procedure produces the matrix
\begin{align}
    A^{\mathrm{final}}_{(n+m)\times(n+m)}&=\begin{bmatrix}
        \bm{\mathsf{A}}+\bm{\mathsf{C}}-\bm{\mathsf{C}}^\intercal+\bm{\mathsf{B}} & & & -\bm{\mathsf{D}}^\intercal-\bm{\mathsf{E}}^\intercal & & & -\bm{\mathsf{C}}^\intercal+\bm{\mathsf{B}} \\
        \bm{\mathsf{D}}+\bm{\mathsf{E}} & & & \bm{\mathsf{F}} & & & \bm{\mathsf{E}} \\
        \bm{\mathsf{C}}+\bm{\mathsf{B}} & & & -\bm{\mathsf{E}}^\intercal & & & \bm{\mathsf{B}} 
    \end{bmatrix} \\ 
    &\!:=\begin{bmatrix}
        \bm{\mathsf{G}}_{m\times m} & -\bm{\mathsf{I}}_{m\times(n-m)}^\intercal & -\bm{\mathsf{J}}_{m\times m}^\intercal \\
        \bm{\mathsf{I}}_{(n-m)\times m} & \bm{\mathsf{H}}_{(n-m)\times(n-m)} & -\bm{\mathsf{K}}_{(n-m)\times m}^\intercal \\
        \bm{\mathsf{J}}_{m\times m} & \bm{\mathsf{K}}_{m\times (n-m)} & \bm{\mathsf{L}}_{m\times m}
    \end{bmatrix},
\end{align}
where now the matrices are 
\begin{equation}
    \begin{split}
        \bm{\mathsf{G}}&=\begin{bmatrix}
            0 & \frac{\mathbb{k}_1\cdot\mathbb{k}_2}{u_{12}} & \cdots & \frac{\mathbb{k}_1\cdot\mathbb{k}_m}{u_{1m}} \\
            \frac{\mathbb{k}_2\cdot\mathbb{k}_1}{u_{21}} & 0 & \cdots & \frac{\mathbb{k}_2\cdot\mathbb{k}_m}{u_{2m}} \\
            \vdots & \vdots & \ddots & \vdots \\
            \frac{\mathbb{k}_m\cdot\mathbb{k}_1}{u_{m1}} & \frac{\mathbb{k}_m\cdot\mathbb{k}_2}{u_{m2}} & \cdots & 0
        \end{bmatrix}, \\
        \bm{\mathsf{H}}&=\begin{bmatrix}
0 & \frac{\mathbb{k}_{m+1}\cdot \mathbb{k}_{m+2}}{u_{(m+1)(m+2)}} & \cdots & \frac{\mathbb{k}_{m+1}\cdot \mathbb{k}_{n}}{u_{(m+1)n}} \\
 \frac{\mathbb{k}_{m+2}\cdot \mathbb{k}_{m+1}}{u_{(m+2)(m+1)}} & 0 & \cdots & \frac{\mathbb{k}_{m+2}\cdot \mathbb{k}_{n}}{u_{(m+2)n}}  \\
\vdots & \vdots & \ddots & \vdots \\
 \frac{\mathbb{k}_{n}\cdot \mathbb{k}_{m+1}}{u_{n(m+1)}} & \frac{\mathbb{k}_{n}\cdot \mathbb{k}_{m+2}}{u_{n(m+2)}}  & \cdots & 0
\end{bmatrix}\,, \\
        \bm{\mathsf{K}}&=\begin{bmatrix}
 \frac{\bbepsilon_{1}\cdot\mathbb{k}_{m+1}}{u_{1(m+1)}} & \frac{\bbepsilon_{1}\cdot\mathbb{k}_{m+2}}{u_{1(m+2)}} & \cdots & \frac{\bbepsilon_{1}\cdot\mathbb{k}_{n}}{u_{1n}} \\
  \frac{\bbepsilon_{2}\cdot\mathbb{k}_{m+1}}{u_{1(m+1)}} & \frac{\bbepsilon_{2}\cdot\mathbb{k}_{m+2}}{u_{2(m+2)}} & \cdots & \frac{\bbepsilon_{2}\cdot\mathbb{k}_n}{u_{2n}} \\
 \vdots & \vdots & \ddots & \vdots \\
 \frac{\bbepsilon_{m}\cdot\mathbb{k}_{m+1}}{u_{m(m+1)}} & \frac{\mathbb{k}_{m}\cdot \bbepsilon_{m+2}}{u_{m(m+2)}} & \cdots & \frac{\bbepsilon_{m}\cdot\mathbb{k}_{n}}{u_{mn}} \\
\end{bmatrix},
    \end{split}
    \qquad
    \begin{split}
        \bm{\mathsf{I}}&=\begin{bmatrix}
 \frac{\mathbb{k}_{m+1}\cdot \mathbb{k}_{1}}{u_{(m+1)1}} & \frac{\mathbb{k}_{m+1}\cdot \mathbb{k}_{2}}{u_{(m+1)2}} & \cdots & \frac{\mathbb{k}_{m+1}\cdot \mathbb{k}_{m}}{u_{(m+1)m}} \\
  \frac{\mathbb{k}_{m+2}\cdot \mathbb{k}_{1}}{u_{(m+2)1}} & \frac{\mathbb{k}_{m+2}\cdot \mathbb{k}_{2}}{u_{(m+2)2}} & \cdots & \frac{\mathbb{k}_{m+2}\cdot \mathbb{k}_{m}}{u_{(m+2)m}} \\
 \vdots & \vdots & \ddots & \vdots \\
 \frac{\mathbb{k}_{n}\cdot \mathbb{k}_{1}}{u_{n1}} & \frac{\mathbb{k}_{n}\cdot \mathbb{k}_{2}}{u_{n2}} & \cdots & \frac{\mathbb{k}_{n}\cdot \mathbb{k}_{m}}{u_{nm}} \\
\end{bmatrix}, \\
        \bm{\mathsf{J}}&=\begin{bmatrix}
 \frac{\hat s_{1'1}}{x_1} & \frac{\bbepsilon_{1}\cdot \mathbb{k}_{2}}{u_{12}} & \cdots & \frac{\bbepsilon_{1}\cdot \mathbb{k}_{m}}{u_{1m}} \\
  \frac{\bbepsilon_{2}\cdot \mathbb{k}_{1}}{u_{21}} & \frac{\hat s_{2'2}}{x_2} & \cdots & \frac{\bbepsilon_{2}\cdot \mathbb{k}_{m}}{u_{2m}} \\
 \vdots & \vdots & \ddots & \vdots \\
 \frac{\bbepsilon_{m}\cdot \mathbb{k}_{1}}{u_{m1}} & \frac{\bbepsilon_{m}\cdot \mathbb{k}_{2}}{u_{m2}} & \cdots & \frac{\hat s_{m'm}}{x_m} \\
\end{bmatrix},\\
        \bm{\mathsf{L}}&=\begin{bmatrix}
0 & \frac{\bbepsilon_{1}\cdot \bbepsilon_{2}}{u_{12}} & \cdots & \frac{\bbepsilon_{1}\cdot \bbepsilon_{m}}{u_{1m}} \\
 \frac{\bbepsilon_{2}\cdot \bbepsilon_{1}}{u_{21}} & 0 & \cdots & \frac{\bbepsilon_{2}\cdot \bbepsilon_{m}}{u_{2m}}  \\
\vdots & \vdots & \ddots & \vdots \\
 \frac{\bbepsilon_{m}\cdot \bbepsilon_{1}}{u_{m1}} & \frac{\bbepsilon_{m}\cdot \bbepsilon_{2}}{u_{m2}}  & \cdots & 0
\end{bmatrix}.
    \end{split}
\end{equation}
The kinematic structure of the matrix $A^{\mathrm{final}}$ is very telling and is urging us to define
\begin{equation}
    \widehat{A}:=\begin{bmatrix}
        \bm{\mathsf{G}} & -\bm{\mathsf{I}}^\intercal \\
        \bm{\mathsf{I}} & \bm{\mathsf{H}}
    \end{bmatrix}, \quad \widehat{B}:=\bm{\mathsf{L}}\,, \quad \widehat{C}:=\begin{bmatrix}
        \bm{\mathsf{J}} & \bm{\mathsf{K}}
    \end{bmatrix},
\end{equation}
so that 
\begin{equation}
    \tilde{\Psi}:=A^{\mathrm{final}}=\begin{bmatrix}
        \widehat{A} & -\widehat{C}^\intercal \\
        \widehat{C} & \widehat{B}
    \end{bmatrix}.
\end{equation}
After imposing the scattering equations in the form of eq.~\eqref{eq:scattering_eq_diag_terms}, the matrix $\tilde{\Psi}$ is precisely the matrix appearing in the upper $(n+m)\times(n+m)$ block in eq.~\eqref{eq:block_compactification}, \ie what we called $\widehat{\Psi}_{2n}$. The matrix $\widehat{A}$ encodes the Mandelstam invariants of the particles, $\widehat{B}$ encodes the polarisations of collinearized particles, and $\widehat{C}$ encodes the interaction between momenta and polarisations.

Since $A^\mathrm{final}$ and $A_{(n+m)\times(n+m)}$ are related by column and row operations, we have that $\Pf'(\tilde{\Psi})=\Pf'(A)$ to leading order in $\tau$, showing that 
\begin{align}
\label{eq:YMS_multi-collinear}
    &A_{n+m}^\mathrm{YMS:scalar}(1,1',\dots,m,m',m+1,\dots,n) \nonumber \\
    &=\left(\prod\limits_{i=1}^m \frac{1}{s_{i i'}}\right)\frac{1}{\mathrm{Vol}(\mathrm{SL}(2,\amsbb{C}))}\int \biggr(\prod \limits_{I=4}^{n} \dd u_I \,\delta(\amsbb{E}_I)\biggr)\,\mathrm{PT}(1,\dots,n)\Pf(N)\Pf'(\tilde{\Psi}) + \dots \nonumber \\ 
    &=\prod_{i=1}^m\frac{1}{s_{ii'}}A_n^{\mathrm{YMS}}(1,\dots,n)+\dots\,,
\end{align}
where it is understood that particles $1$ to $m$ are gluons, and particles $m+1$ to $n$ are scalars. The last equality follows from the formula for the compactification of Yang-Mills and the Pfaffian of a block matrix -- see the discussion surrounding eq.~\eqref{eq:block_compactification}. The case $m=n$ reproduces the scaffolding residue studied in section~\ref{sec:YMS}.

\subsection{Color-dressed amplitudes}

A logical follow-up is to consider the multi-collinear limit of full color-dressed amplitudes and derive the analogue of eq.~\eqref{eq:YMS_multi-collinear}.

The full color-dressed amplitude of Yang-Mills with gauge group $U(N)$ is built of out color-ordered amplitudes in the following way \cite{ManganoReview}:
\begin{equation}
\label{eq:full_color_dressed_amplitude}
\mathcal{A}^{\mathrm{YMS:scalar}}_{n}= \sum_{\omega\,\in \,S_n/ \amsbb{Z}_n} \Tr\Big(T^{\dsa_{\omega(1)}}\cdots T^{\dsa_{\omega(n)}} \Big) A^{\mathrm{YMS:scalar}}_{n}(\omega)\,,
\end{equation}
where $T^\dsa$ are a fixed choice of generators of $U(N)$ and $A_n^{\mathrm{YMS:scalar}}(\omega)$ is the non-canonically-ordered partial amplitude
\begin{equation}
\label{eq:ordered_partial_amplitude}
    A_n^{\mathrm{YMS:scalar}}(\omega)=\frac{1}{\mathrm{Vol}(\mathrm{SL}(2,\amsbb C))}\int\prod_{i=1}^n \mathfrak{d}u_i \,\mathrm{PT}(\omega(1,\dots,n))\,\Pf{X}\Pf'{A}\,,
\end{equation}
which is a generalisation to eq.~\eqref{eq:YMS_amplitude}. Eq.~\eqref{eq:full_color_dressed_amplitude} is known as a color decomposition of the amplitude. We note that the l.h.s. depends on, for each particle: the momentum $k_i$; the (possibly vanishing) polarisation vector $\epsilon_i$; and the color $\dsa_i$, which must take values between $1$ and $N^2$.

We again consider the (color-dressed) amplitude of $n+m$ particles using a labelling tailored to the limit $(1,1',\dots,m,m',m+1,\dots,n)$. Under the multi-collinear limit $s_{ii'}\to 0$ parametrised by $\tau$, with $1\leq i\leq m$, the most singular term of eq.~\eqref{eq:full_color_dressed_amplitude} contains the terms of the sum where ordering $\omega\in S_{n+m}/\amsbb{Z}_{n+m}$ leaves the primed and unprimed particles adjacent. Any other ordering is sub-leading in $\tau$ because the Parke-Taylor factor in \eqref{eq:ordered_partial_amplitude} will not contribute enough factors of $\tau$. In light of this, the permutations in the most singular term are really $\omega\in S_n/\amsbb{Z}_n$, acting on elements of the set $\{(1,1'), \dots, (m,m'), m+1, \dots, n\}$, with a possible transposition among the $m$ pairs, for a total of $2^m(n-1)!$ terms. 

We would like to show that in the multi-collinear limit, the color-dressed amplitude has the pole expansion
\begin{equation}
\label{eq:multi-collinear_color_dressed}
    \mathcal{A}_{n+m}^{\mathrm{YMS:scalar}}=\prod_{i=1}^m\frac{1}{s_{ii'}}\left(\prod_{j=1}^m f_{jj'j^\star}\mathcal{A}_n^{\mathrm{YMS}}\right)+\dots\,,
\end{equation}
where
\begin{equation}
    f_{jj'j^\star}:=f_{\dsa_{j}\dsa_{j'}\dsa_{j^\star}}\,,
\end{equation}
are structure constants of $\mathfrak{u}(N)$ and
\begin{equation}
   \mathcal{A}^{\mathrm{YMS}}_{n}= \sum_{\omega\,\in \,S_n/ \amsbb{Z}_n} \Tr\Big(T^{\dsa_{\omega(1^\star)}}\cdots T^{\dsa_{\omega(m^\star)}}T^{\dsa_{\omega(m+1)}}\cdots T^{\dsa_{\omega(n)}} \Big) A^{\mathrm{YMS}}_{n}(\omega)\,.
\end{equation}
Let us take a moment to explain the notation. On the l.h.s.\@ of eq.~\eqref{eq:multi-collinear_color_dressed}, particles $\{1,1',\dots,m,m',m+1,\dots,n\}$ are scattered. After the multi-collinear limit, we are left with particles $\{1^\star,\dots m^\star,m+1,\dots,n\}$, where particles $j,j'$ have fused to give $j^\star$. Previously, we employed an abuse of notation by dropping the stars, but now we simply cannot, as the color structure on the r.h.s. of eq.~\eqref{eq:full_color_dressed_amplitude} depends on the color of particles $j$ and $j'$, and the newly fused $j^\star$. Furthermore, it is understood that all combinations of colors of particles $j^\star$ for $1\leq j\leq m$ are summed over. To perform this sum, $a_{j^\star}$ are Einstein summation indices, running from $1$ to $N^2$. Let us emphasize that $a_{j^\star}$ for $1\leq j\leq m$ are summation indices, while $a_{j}$ and $a_{j'}$ for $1\leq j\leq n$ are color data provided for the particles.

\textbf{Example: $m=n=3$.} To elucidate this notation and formula let us consider the scaffolding residue of the color-dressed amplitude $\mathcal{A}_{3+3}^{\mathrm{YMS}}$. The sum over permutations in \eqref{eq:full_color_dressed_amplitude} yields $2^3(3-1)!=16$ terms, which can be grouped and simplified using the linearity of the trace to obtain
\begin{align}
\label{eq:example_m=n=3}
\begin{split}
    \mathcal{A}_{3+3}^{\mathrm{YMS}}=f_{11'1^\star}f_{22'2^\star}f_{33'3^\star} \times \Bigg(&\Tr(T^{\dsa_{1^\star}}T^{\dsa_{2^\star}}T^{\dsa_{3^\star}})A_{3+3}^{\mathrm{YMS}}(11'22'33')\\
    &+\Tr(T^{\dsa_{1^\star}}T^{\dsa_{3^\star}}T^{\dsa_{2^\star}})A_{3+3}^{\mathrm{YMS}}(11'33'22')\Bigg)+ \dots\,.
\end{split}
\end{align}
The scaffolding limit $s_{ii'}\to 0$ passes through the structure constants and traces to the kinematic partial amplitudes. Although we didn't explicitly study the scaffolding residue of a non-canonically-ordered partial amplitude, it trivially generalises in the way one expects, that the $i,i'$ particles fuse together and we get a pole at the on-shell momenta. Thus, the bracketed term of \eqref{eq:example_m=n=3} is precisely the color-dressed amplitude $\mathcal{A}_3^{\mathrm{YMS}}$ of three particles, summed over all possible combinations of colors of the particles (multiplied by the corresponding color structure constants), \ie
\begin{equation}
    \mathcal{A}_{3+3}^{\mathrm{YMS}}=\left(\prod_{i=1}^3 \frac{1}{s_{ii'}}\right)f_{11'1^\star}f_{22'2^\star}f_{33'3^\star}\mathcal{A}_3^\mathrm{YMS}\,.
\end{equation}

\noindent \textbf{Proof of eq.~\eqref{eq:multi-collinear_color_dressed}.} We make the observation that the multi-collinear limit can be taken sequentially, \ie it is equivalent to first taking $s_{11'}\to 0$, then $s_{22'}\to0$, and so on until $s_{mm'}\to0$. The order in which the limits are taken can be permuted. Without loss of generality, it is sufficient to show the single collinear limit $s_{11'}\to 0$ produces the correct pole for a color-dressed amplitude. 

For simplicity, let the group of circular permutations on $n$ elements be denoted $P_n:=S_{n}/\amsbb{Z}_{n}$. To avoid writing long products of permuted generators we denote
\begin{equation}
    T_\omega(1,\dots,n):=T^{\dsa_{\omega(1)}}\cdots T^{\dsa_{\omega(n)}}\,.
\end{equation}

We begin with the color-dressed amplitude of $n+1$ particles labelled $(1,1',2,\dots,n)$:
\begin{equation}
    \mathcal{A}^{\mathrm{YMS:scalar}}_{n+1}= \sum_{\omega\,\in \,P_{n+1}} \Tr\Big(T_\omega(1,1',2\dots,n) \Big) A^{\mathrm{YMS:scalar}}_{n}(\omega)\,.
\end{equation}
Recall that only permutations which keep the $1,1'$ particles adjacent contribute to the most singular term in the collinear limit. Keeping only these terms, we can also equivalently write the sum over circular permutations on the remaining terms as follows:
\begin{align}
    \mathcal{A}_{n+1}^{\mathrm{YMS:scalar}}=\sum_{\sigma\,\in \,S_{n-1}}&\Tr\Big(T^{\dsa_1}T^{\dsa_{1'}}T_\sigma(2,\dots,n)\Big)A^{\mathrm{YMS:scalar}}_{n+1}(1,1',\sigma) \nonumber \\
    &-\Tr\Big(T^{\dsa_{1'}}T^{\dsa_{1}}T_\sigma(2,\dots,n)\Big)A^{\mathrm{YMS:scalar}}_{n+1}(1,1',\sigma)+\dots \,,
\end{align}
the negative sign arising from the Parke-Taylor factor in the partial amplitude. We can collect the two terms together to get
\begin{equation}
    \mathcal{A}_{n+1}^{\mathrm{YMS:scalar}}=f_{11'1^\star}\!\!\!\sum_{\sigma\,\in \,S_{n-1}}\Tr\Big(T^{\dsa_{1^\star}}T_\sigma(2,\dots,n)\Big)A_{n+1}^{\mathrm{YMS:scalar}}(1,1',\sigma)+\dots\,.
\end{equation}
Now taking the collinear limit $s_{11'}\to 0$, which passes through the color structure, we have
\begin{equation}
    \mathcal{A}_{n+1}^{\mathrm{YMS:scalar}}=\frac{1}{s_{11'}}f_{11'1^\star}\!\!\!\sum_{\sigma\,\in \,S_{n-1}}\Tr\Big(T^{\dsa_{1^\star}}T_\sigma(2,\dots,n)\Big)A_{n+1}^{\mathrm{YMS}}(1^\star,\sigma)+\dots=\frac{1}{s_{11'}}f_{11'1^\star}\mathcal{A}^{\mathrm{YMS}}_{n}+\dots\,.
\end{equation}
Eq.~\eqref{eq:multi-collinear_color_dressed} now follows from taking additional collinear limits on other particles $i,i'$, one at a time. $\crule[black]{2.5mm}{2.5mm}$

\section{Discussions}

In this Winter School Project we revisited the computation of collinear limits of various amplitudes using their CHY formulation. The main case we considered was the so-called ``scaffolding" residues of amplitude of scalars in the Yang-Mills-Scalar (YMS) theory. While the scaffolding residue was introduced and studied in ref.~\cite{Arkani-Hamed:2023jry}, here we computed it using the corresponding CHY formulations. At first sight, obtaining an $n$-gluon amplitude CHY formula from that of a $2n$-scalar amplitude in YMS seems to be a daunting task. The reason is that the scalar theory is built using a very simple CHY integrand, the pfaffian of $A$, while that of gluons uses one of the most complicated, the pfaffian of $\Psi$. Not only were we able to show how one naturally emerges from the other, but we showed how one of the most mysterious features of the $\Psi$ matrix, the $C_{ii}$ elements
\begin{equation}
    C_{ii} = -\sum_{j=1}^n \frac{\epsilon_i\cdot k_j}{u_{ij}}\,,
\end{equation}
arises from the scattering equations. 

Some next steps include understanding how even more complicated matrices in the CHY formulation can arise from simpler ones by similar scaffolding procedure. For example, the matrix $\Pi$ in the CHY formula of Einstein-Yang-Mills amplitude has a very rich structure and it would be fascinating to see it emerge from a simple matrix.

\section*{Acknowledgements}

This research was done as part of the Perimeter Scholars International (PSI) program at Perimeter Institute and ZJ, JS, and YY would like to thank Perimeter Institute for its support.
This research was supported in part by a grant from the Gluskin Sheff/Onex Freeman Dyson Chair in Theoretical Physics and by Perimeter Institute. Research at Perimeter Institute is supported in part by the Government of Canada through the Department of Innovation, Science and Economic Development Canada and by the Province of Ontario through the Ministry of Colleges and Universities.

\bibliographystyle{JHEP}
\bibliography{references.bib}

\end{document}